\documentclass[fleqn,usenatbib]{mnras}
\usepackage{blindtext}
\usepackage{newtxtext,newtxmath}
\usepackage[T1]{fontenc}

\usepackage{graphicx}	% Including figure files
\usepackage{amsmath}	% Advanced maths commands
\usepackage{multicol,tabularx,capt-of}
\usepackage{hhline}
\usepackage{multirow}
\title[A new precursor for solar cycle prediction]{Discovery of a relation between the decay rate of the Sun's magnetic dipole and the growth rate of the following sunspot cycle: a new precursor for solar cycle prediction}

\author[P. Jaswal et al.]{
Priyansh Jaswal,$^{1}$
Chitradeep Saha,$^{1}$
Dibyendu Nandy$^{1,2}$\thanks{E-mail: dnandi@iiserkol.ac.in}
\\
$^{1}$Center of Excellence in Space Sciences India, Indian Institute of Science Education and Research Kolkata, Mohanpur 741246, West Bengal, India\\
$^{2}$Department of Physical Sciences, Indian Institute of Science Education and Research Kolkata, Mohanpur 741246, West Bengal, India\\
}

\date{Accepted XXX. Received YYY; in original form ZZZ}

\pubyear{2023}

\begin{document}
\label{firstpage}
\pagerange{\pageref{firstpage}--\pageref{lastpage}}
\maketitle

% Abstract of the paper
\begin{abstract}
Sunspots have been observed for over four centuries and the magnetic nature of sunspot cycles has been known for about a century; however, some of its underlying physics still remain elusive. It is known that the solar magnetic cycle involves a recycling of magnetic flux between the poloidal and toroidal components of the magnetic field, that manifests as the solar dipole and sunspots, respectively. Here we report the discovery of a new relationship between the rise rate of the sunspot cycle and the decay rate of the solar (axial) dipole moment. This provides an extension to the Waldmeier effect in sunspot cycles and points to the existence of a causal connection between the aforementioned physical quantities, which can be succinctly stated as \textit{ the decay rate of the Sun's dipole moment is related to the rate of rise of the following sunspot cycle}. We demonstrate how one may take advantage of this new relationship to predict the timing of the sunspot cycle. Our analysis indicates solar cycle 25 is expected to be a weak-moderate cycle, peaking in \(2024.00_{-0.49}^{+0.68} \).

% It should be a single paragraph not more than 250 words (200 words for Letters).
% No references should appear in the abstract.
\end{abstract}

% Select between one and six entries from the list of approved keywords.
% Don't make up new ones.
\begin{keywords}
Sun: activity -- Sun: magnetic fields -- Sun: interior
\end{keywords}

%%%%%%%%%%%%%%%%%%%%%%%%%%%%%%%%%%%%%%%%%%%%%%%%%%

%%%%%%%%%%%%%%%%% BODY OF PAPER %%%%%%%%%%%%%%%%%%

\section{Introduction}

Our host star, the Sun, is a dynamic star whose magnetic activity varies across a wide range of timescales spanning from minutes to millennia and beyond \citep{Usoskin_2023_SolPhys}.  The most prominent signature of this variability is captured by the waxing and waning of sunspots -- dark, magnetized patches on the Sun's surface -- that repeats almost every 11 years, known as the sunspot cycle. Sunspot cycles exhibit significant fluctuations in both amplitude and duration that occasionally result in  extreme activity phases like solar grand minima and grand maxima \citep{Passos_2014_AandA, Hazra_2019_MNRAS, Saha_2022_MNRASL, Dash_2023_MNRAS}. The Sun's  dynamic activity output  influences the entirety of the heliosphere including our home planet, the Earth, by shaping its space environmental conditions and determining the habitability \citep{Shcrijver_2015_AdvSpRes, Nandy_2021_PEPS, NANDY_2023_JASTP}. Therefore, developing accurate predictive capabilities pertaining to the long-term solar activity is crucial in planning future space missions and safeguarding space-reliant technologies \citep{Petrovay_2020_SolPhys, Nandy_2021_SolPhys, Bhowmik_2023_ISSI}.

Stripped down to its fundamental essence, the magnetic activities of the Sun originate in its deep interior, wherein, a magnetohydrodynamic dynamo action generates and recycles the Sun's large-scale magnetic fields \citep{Nandy_2002_Science, Chatterjee_2004_AandA, Charbonneau_2020_LivRev}. The emergence of magnetic flux on the solar surface and its poleward migration under various flux-transport processes like supergranular diffusion, meridional circulation, etc. contribute to the gradual build up of global solar axial dipole moment (hereafter, dipole moment) \citep{Dasi_2010_AandA, Pal_2023_ApJ, Hazra_2023_ISSI}. It is evident from observations that the mean latitude of sunspot emergence drifts towards the equator with the progress of sunspot cycles \citep{Li_2003_SolPhys, CameronandSchussler_2007_ApJ, Solanki_2008_AandA, Owens_2011_GRL, Mandal_2017_ApJ}, thereby facilitating cross-equatorial diffusion of magnetic fluxes and their cancellation across the equatorial region.

\begin{figure*}
    \centering
    \includegraphics[width = 0.9\textwidth]{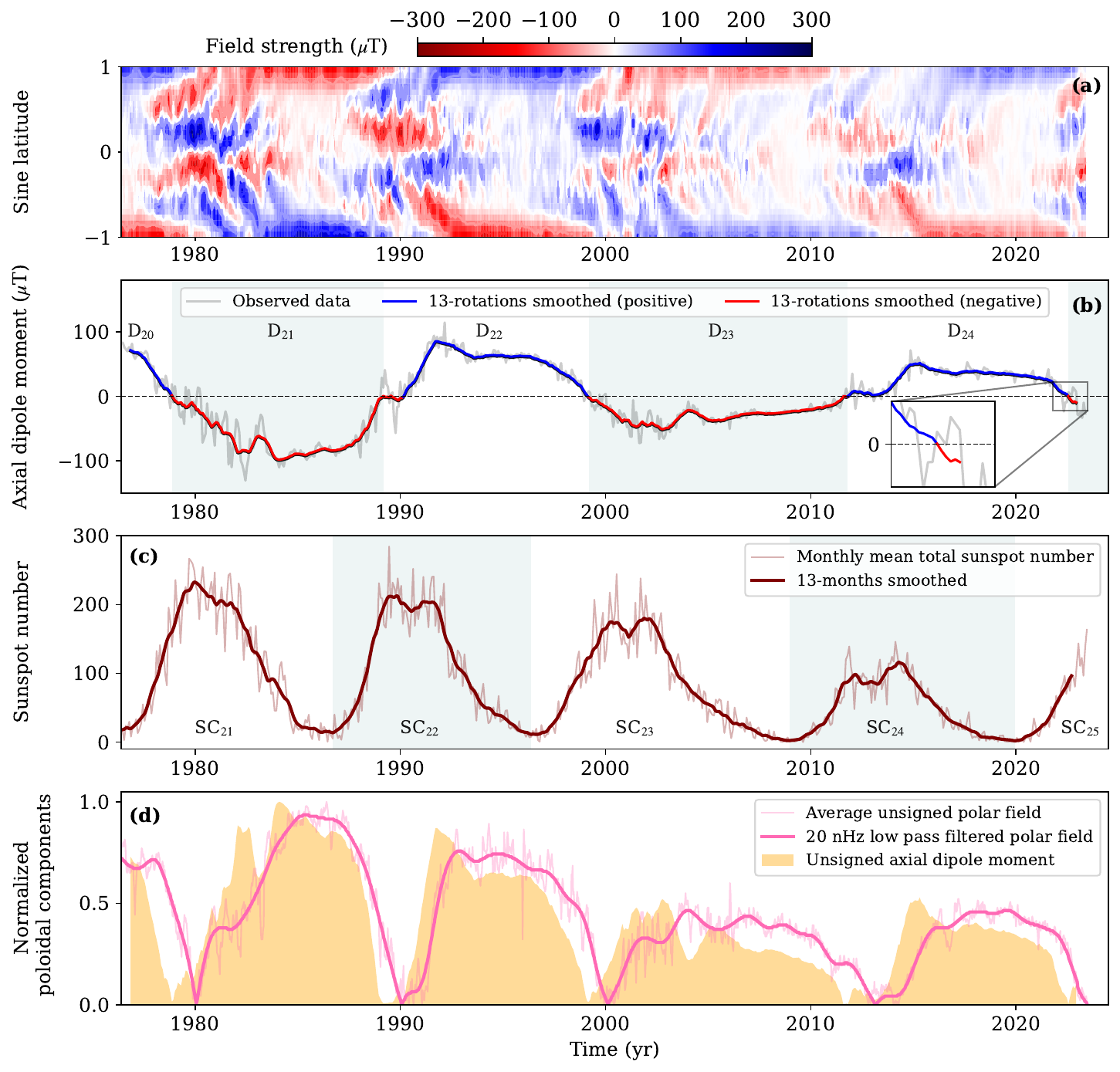}
    \caption{Panel (a):  magnetic butterfly diagram showing the longitudinally averaged line-of-sight solar photospheric magnetic field since May 1976 to May 2023 (i.e., Carrington Rotation number 1642-2271) gleaned from the Wilcox Solar Observatory (WSO) synoptic charts. Panel (b): the grey curve in the background depicts the evolution of solar axial dipole moment cycles for the above mentioned period. Blue and red curves in the foreground represent 13-rotations smoothed (uniform running average) dipole moment denoting its positive and negative global polarity, respectively. Alternately shaded intervals in the background delineate consecutive dipole moment cycles with the cycle numbers D\(_{20-24}\) labelled on the plot. The inset plot zooms into the tail end of dipole moment time series emphasizing the latest polarity reversal in solar dipole moment -- from positive (in blue) to negative (in red) -- that occurred during July 2022. This reversal in polarity heralds the approaching arrival of the peak of sunspot cycle 25. Panel (c): monthly mean total sunspot number time series (in the background) and its 13-months uniform running average (in the foreground) for the aforementioned period, i.e. since sunspot cycle 21 to present. Alternately shaded intervals in the background depicts individual sunspot cycles with the cycle numbers SC\(_{21-25}\) labelled on the plot. Sunspot number data is obtained from WDC-SILSO, Royal Observatory of Belgium, Brussels. Panel (d): juxtaposition of two normalized time series -- namely, the unsigned axial dipole moment (shaded in yellow) and hemispherically averaged unsigned polar field (in pink), both observed by WSO -- depicts a finite time/phase lag in the latter with respect to the former one. }
    \label{fig:Figure 1}
\end{figure*}

\begin{figure*}
    \centering
    \includegraphics[width = 0.9\textwidth]{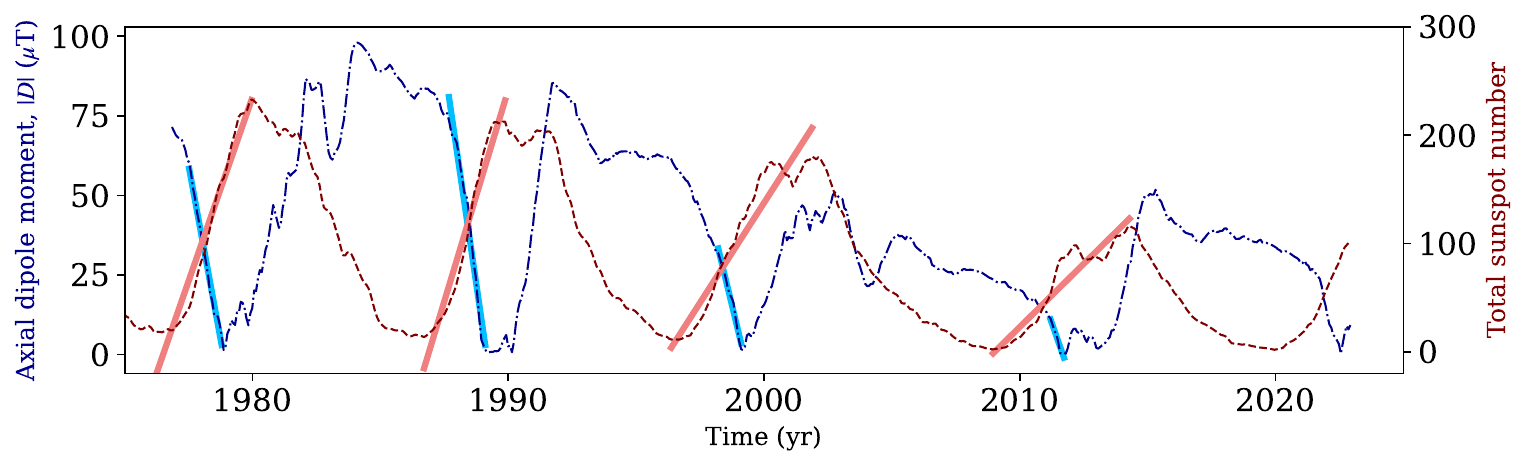}
    \caption{Evolution of 13-months smoothed monthly total sunspot number since sunspot cycle 21 (in red-dashed curve) and corresponding unsigned dipole moment, \(\vert D\vert\) (in blue dash-dotted curve). In our analyses, the slopes of the linearly fitted blue and red solid lines determine the decay rate of unsigned dipole moment, \(r_{\text{DM}}\), and the rise rate of sunspot cycles, \(r_{\text{SSN}}\), respectively.}
    \label{fig:Figure 2}
\end{figure*}

\begin{table*}
    \centering    
    \caption{Calculated rise rate of previous four sunspot cycles \(\text{SC}_{21-24}\) and the decay rate of their precursor dipole moment cycles \(\text{D}_{20-23}\) are tabulated. Initial and final time of each interval, as considered in our analyses, are also reported (in year). Corresponding Carrington Rotation (CR) numbers are mentioned in parentheses.}
    \begin{tabular}{c c c c c c c c c}
         \hline \hline
          \multirow{1}{*}{Sunspot cycle}&\multirow{1}{*}{Dipole moment cycle}& \multicolumn{3}{c}{Decay of precursor dipole cycle $\vert$D$_{n-1}\vert$} & & \multicolumn{3}{c}{Rise of sunspot cycle SC$_{n}$} \\
          \cline{3-5} \cline{7-9}
          SC$_{n}$& D$_{n-1}$ & Initial time  & Final time  & Decay rate, $r_{\text{DM}}$  & &  Initial time  & Final time  & Rise rate, $r_{\text{SSN}}$ \\
          &  & [yr (CR)]  & [yr (CR)] &  [$\mu$T yr\(^{-1}\)] & &  [yr]  &  [yr] & [yr\(^{-1}\)]\\
          \hline
          SC$_{21}$&D$_{20}$ & 1977.60 (CR 1658) & 1978.87 (CR 1675) & 43.5917 & &  1976.21 & 1979.96 & 68.0175\\
          SC$_{22}$&D$_{21}$ & 1987.75 (CR 1794) & 1989.17 (CR 1813) & 54.9517 & &  1986.71 & 1989.87 & 78.0974\\
          SC$_{23}$&D$_{22}$ & 1998.28 (CR 1935) & 1999.18 (CR 1947) & 33.2563 & &  1996.34 & 2001.87 & 36.8719\\
          SC$_{24}$&D$_{23}$ & 2011.28 (CR 2109) & 2011.80 (CR 2116) & 22.8997 & &  2008.96 & 2014.23 & 23.3260\\
          SC$_{25}$&D$_{24}$ & 2021.28 (CR 2243) & 2022.55 (CR 2260) & 26.0578 & &  2019.96 & 2022.87 & --\\
          \hline
    \end{tabular}
    \label{tab:tab1}
\end{table*}

Recently, \cite{Ijima_2017_AandA} demonstrated that the emergence of new sunspots during the decaying phase of a sunspot cycle do not have considerable influence on the polar field build up. In fact, earlier studies have detected plateau-like intervals in the dipole moment time series -- showing no substantial changes in its magnitude for an extended duration of multiple years -- during the descending phase of sunspot cycles 21 to 24 \citep{Schrijver_2008_SoPh, Ijima_2017_AandA}. On the other hand, meridional circulation, turbulent diffusion and turbulent magnetic pumping are believed to work in tandem to advect poloidal fields accumulated in the polar caps down into the base of solar convection zone (SCZ), where strong radial and latitudinal shear induct toroidal field that acts as a seed for the next sunspot cycle  \citep{Yeates_2008_ApJ,Jaramillo_2009_ApJ, Cameron_2015_Science}. Generation of toroidal field in SCZ consumes the poloidal field of previous cycle.
As a matter of fact, the solar dipole moment comes out of the plateau-like phase and starts decaying abruptly with almost a uniform rate. Besides, the toroidal fields produced at the base of SCZ become buoyantly unstable, rise up through the convection zone in the form of magnetic flux tubes and penetrates the solar surface -- thereby producing sunspots of the new cycle. Decay and dispersal of these new sets of sunspots eventually lead to a growth in the Sun's poloidal field, but with opposite polarity as compared to the previous cycle (see Fig.\ref{fig:Figure 1}, panel (a)).

This sequence of events indicates the existence of a causal connection between the decay of solar polar fields and dipole moment, and the rise of the following sunspot cycle. In fact it is widely known that steeply rising sunspot cycles peak to higher amplitudes and vice versa -- known as the Waldmeier effect \citep{Waldmeier_1935_Astronomische}. \cite{Kumar_2021_ApJ} found correlation between the decay rate of polar fields and the amplitude of the subsequent sunspot cycle across individual hemispheres of the Sun. However, it is to be noted that the decay of high-latitude polar field is almost concurrent with the ascent of the following sunspot cycle, leading to a narrow temporal window for solar cycle prediction (see, Appendix \ref{sec: Appendix}). In this context, the dipole moment of the Sun has the potential to become a better precursor compared to the high-latitude polar field, where the former leads the latter by about a year as evidenced in observational data (see Fig.\ref{fig:Figure 1}, panel (d)).  \cite{Petrovay_2020_SolPhys} argued this time lag to originate from the delay induced by the poleward transport of low- and mid-latitude magnetic fields -- during the formation of high-latitude polar fields. 

In this work, we investigate the relationship between the declining phase of the axial dipole moment associated with the solar cycle and the rise rate of the following sunspot cycle. We find a compelling relationship between the two. We argue that this is theoretically expected and points to a causal connection between the flux transport dynamics mediated dispersal of active region flux during the rise of a sunspot cycle and the cancellation of the polar field of the previous cycle. Furthermore, we demonstrate how this new relationship can be utilized to predict the future sunspot cycle, especially the timing of its peak which is a challenging task. Our results also support the Babcock-Leighton paradigm of the sunspot cycle which proposes that the decay and dispersal of the flux of tilted bipolar sunspot pairs mediated via surface flux transport processes is the primary mechanism for solar poloidal field's creation.

\section{Methods and Results}

\begin{figure}
    \centering
    \includegraphics[width = 0.44\textwidth]{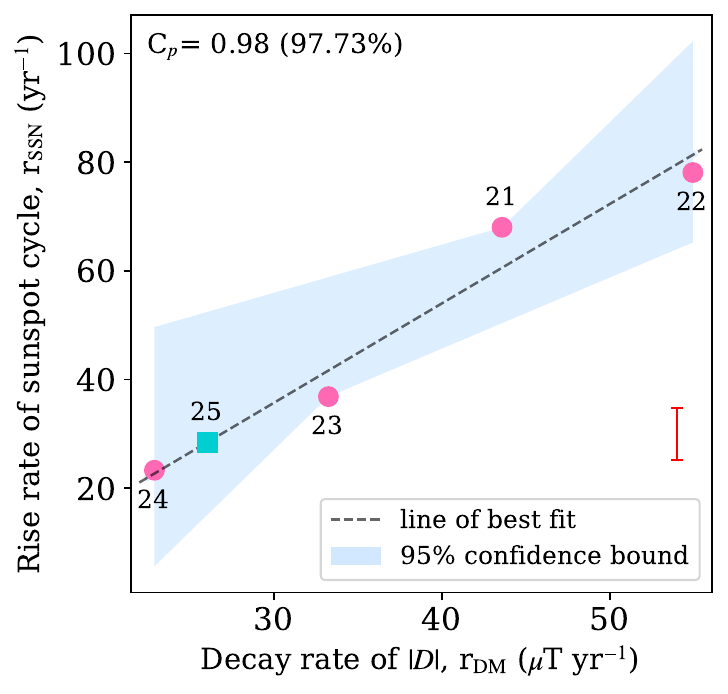}
    \caption{Evidence of a strong correlation (Pearson's r = 0.98 with confidence level of 97.73\%) between the decay rate of unsigned dipole moment, \(r_{\text{DM}}\), and the rise rate of the following sunspot cycle, \(r_{\text{SSN}}\). The black-dashed line denotes the best-fitted curve, while the shaded region in the background marks the corresponding 2\(\sigma\) confidence bound as obtained from linear regression. The error bar represents the typical magnitude of root-mean-squared error (RMSE) associated with this regression model, considering no other statistical uncertainties. Sunspot cycle numbers (21-25) are mentioned adjacent to their respective data points in the plot. The predicted rise rate of sunspot cycle 25 using this model is 28.5\(\pm\)4.7 sunspots per year, as denoted by the blue square.}
    \label{fig:Figure 3}
\end{figure}

We make use of total sunspot number database maintained by the SIDC-SILSO and the solar synoptic charts recorded at the Wilcox Solar Observatory (WSO), covering the information of photospheric solar magnetic activity since 1976 to 2023. For a given synoptic chart corresponding to a particular Carrington Rotation number associated with time $t$, global axial dipole moment of the Sun, \(D\), at that instant can be formulated as, \citep[see][]{Petrovay_2020_SolPhys},

\begin{equation}\label{eq:q1}
    D(t) = \frac{3}{2} \int^{\pi}_{0} \overline{B}(\theta, t) \cos{\theta} \sin{\theta}\ d\theta,
\end{equation}

\noindent
where, \(\overline{B}\) represents azimuthally averaged radial magnetic field of the Sun at colatitude $\theta$.

\begin{figure}
    \centering
    \includegraphics[width = 0.44\textwidth]{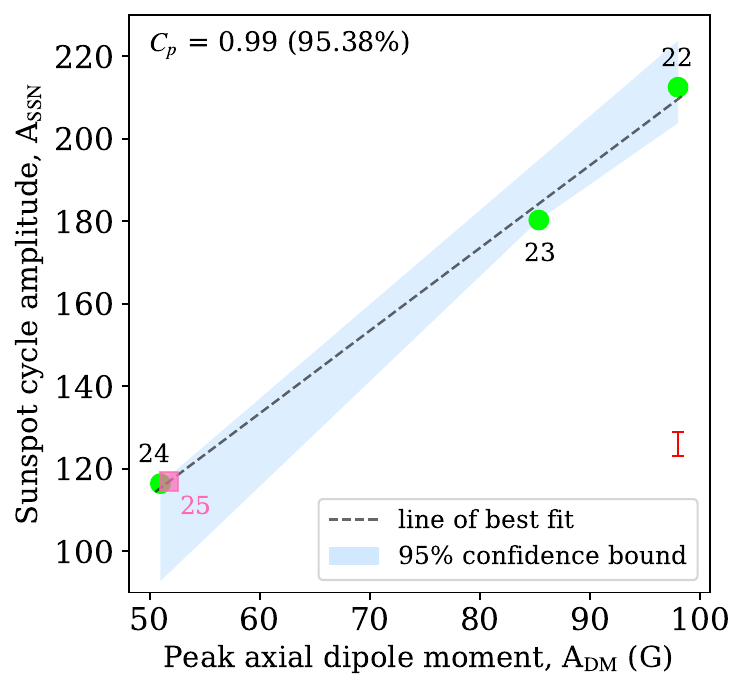}
    \caption{Observed amplitude, \(A_{\text{SSN}}\), of sunspot cycles 22-24 exhibit strong correlation (Pearson's r = 0.99 with 95.38\% confidence level) with the amplitude of preceding unsigned axial dipole moment cycles as observed by WSO, \(A_{\text{DM}}\). Sunspot cycle numbers (22-25) are mentioned adjacent to their respective data points in the plot. Based on the best-fit linear regression model (in black-dashed line) and the observed rate of decay of the preceding \(\vert D\vert\) cycle, the predicted amplitude of sunspot cycle 25 is estimated to be \(116.91\pm2.89\), as denoted by the pink square. The error bar represents the typical magnitude of RMSE associated with the regression model and assuming there are no other statistical uncertainties.}
    \label{fig:Figure 4}
\end{figure}

In the rising phase of a sunspot cycle the number of sunspots surges, accompanied by a fall in the magnitude of solar dipole moment until the latter reverses its global polarity (see Fig. \ref{fig:Figure 1}, panels (b)-(c)). This observation falls in line with the previously mentioned dynamo mechanism pertaining to the cyclic generation of poloidal and toroidal components of the Sun's large-scale magnetic field. Observations show that the polarity reversal of dipole moment precedes the occurrence of sunspot cycle peak by around a year. We hereby report the latest  reversal in polarity of the solar dipole moment to have already occurred almost a year ago, during July 2022 --  which anticipates an imminent cycle maximum of the ongoing sunspot cycle 25.

Since, the growth of a sunspot cycle (say, \(n\)) devours the precursor dipole moment of cycle \((n-1)\), one would expect the time rates of these two physical processes to be in causal correlation with each other. To investigate this, we analyze the time series of the past four sunspot cycles (SC\(_{21-24}\)) and their corresponding precursor dipole moment cycles (D\(_{20-23}\)) by implementing linear regression over their growth and declining phases, respectively (Ref. Fig.\ref{fig:Figure 1} caption for the definitions of SC\(_{21-24}\) and D\(_{20-23}\)). We define, the growth phase of the sunspot cycle as the interval during which the sunspot numbers rise from the cycle minimum to the cycle maximum with the rate, \(r_{\text{SSN}}\). On the other hand, we take a semi-analytical approach (prescribed in Appendix \ref{sec: Appendix}) to determine the decay intervals of individual dipole moment cycles, based on which we estimate their rate of decay, \(r_{\text{DM}}\). We find these two dynamical quantities, namely \(r_{\text{SSN}}\) and \(r_{\text{DM}}\) strongly correlate with each other (Pearson's \(r=0.98\) with 97.73\% confidence level), as described in Fig. \ref{fig:Figure 3}, and the correlation can be expressed as follows,
\begin{equation}
    r_{\text{SSN}} = 1.83 \times r_{\text{DM}} - 19.17
    \label{Eq:Eq.2}
\end{equation}

\noindent
A further investigation of a similar relation as in Eq.\ref{Eq:Eq.2} using the decay rate of WSO average polar field instead of dipole moment demonstrates a positive correlation but with poor statistical significance. Utilizing the observed rate of decay of dipole moment cycle \(D_{24}\) (i.e., \(\sim26.1\ \mu\)T yr\(^{-1}\)) in the empirical relationship prescribed above we estimate the rate of rise of the ongoing sunspot cycle 25 to be \(28.5\pm4.7\) sunspots per year -- which is higher than that of the previous sunspot cycle 24 but lower than cycle 23 (see Table \ref{tab:tab1}). We note that the outcome of the aforementioned regression is sensitive to the choice of initial epoch in the decay interval of dipole moment cycles and we discuss more on this in Appendix \ref{sec: Appendix}.

Now we demonstrate how an amalgamation of this prior knowledge on the rise rate of a sunspot cycle, and its amplitude predicted by other independent means can be extended to forecasting the time of occurrence of its peak. Earlier studies have found that the magnitude of solar polar field and dipole moment at the sunspot cycle minimum significantly correlate with the strength of the subsequent sunspot cycle \citep{Schatten_1978_GRL, Yeates_2008_ApJ, Jiang_2018_ApJ}. Fig. \ref{fig:Figure 4} depicts that even the amplitude of the dipole moment, \(A_{\text{DM}}\), has a significant correlation with the subsequent sunspot cycle amplitude, \(A_{\text{SSN}}\), which can be expressed in the form of the following independent relationship,

\begin{equation}
    A_{\text{SSN}} = 2.00 \times A_{\text{DM}} + 13.16 
    \label{Eq:Eq.3}
\end{equation}

\noindent
Substituting \(A_{\text{DM}} = 51.75  \ \mu\)T (i.e., the observed amplitude of dipole cycle D\(_{24}\)) in equation (\ref{Eq:Eq.3}), we estimate the strength of the imminent sunspot cycle 25 maximum to be \(116.91\pm2.89\) denoting a weak-moderate cycle similar to or slightly stronger than cycle 24. 

We mark the sunspot cycle minimum during December 2019 (say, \(t_{25}^{i}\)) with a monthly mean amplitude of 1.8 (say, \(A_{25}^{i}\)) as the beginning of the ongoing sunspot cycle 25. 
Ascribing a uniform average rise rate to this cycle (i.e., \(r_{25} = 28.5 \pm 4.7\) sunspots per year) as estimated from equation (\ref{Eq:Eq.2}) and considering its amplitude (i.e., \(A_{25}^{f}\) \(=116.91 \pm 2.89\) predicted from equation (\ref{Eq:Eq.3}), we forecast the time of occurrence of the peak of sunspot cycle 25, \(t_{25}^{f}\) to be,

 \begin{equation}
     t_{25}^{f} = t_{25}^{i} + \dfrac{A_{25}^{f} - A_{25}^{i}}{r_{25}} = 2024.00_{-0.49}^{+0.68}
 \end{equation}

Note that in the calculation of the range of possibilities of the expected peak timing we consider only the root-mean-squared error, and no other statistical uncertainties.

\section{Conclusions}
Analyzing long-term observation of solar photospheric magnetic activity for the past four sunspot cycles, we discover a compelling correlation between the decay rate of solar dipole moment and the rise rate of following sunspot cycle. We have explained how this correlation emerges out of a causal connection between the emergence and surface flux transport of new tilted bipolar sunspot pairs (cause) and the decay and reversal of the previous cycle's poloidal field (effect). Given that this causal connection is intimately related to the Babcock-Leighton mechanism for solar polar field generation our work provides independent confirmation that this mechanism is an integral part of the solar dynamo. 

The rise rate of a sunspot cycle (say, cycle $n$) is known to be related to the eventual peak of that sunspot cycle $(n)$ -- a relationship known as the Waldmeier effect. Our work establishes an extension of this Waldmeier effect which can be succinctly stated as: the rate of decay of the Sun's axial dipole moment of cycle $(n - 1)$ is related to the rate of rise, and consequently, the eventual strength of the following sunspot cycle (i.e., cycle $n$).

Additionally, we formulate a semi-analytical framework to determine the decay time interval in dipole moment. It is worth noting that the evolution of the WSO dipole moment precedes that of the average solar polar field by nearly a year, which significantly extends the prediction window for the dynamics of the upcoming sunspot cycle with improved accuracy. The existence of such a strong correlation, in fact, enables one to forecast the timing of a sunspot cycle's peak once the amplitude of that cycle is independently anticipated. For example, we show that the ongoing sunspot cycle is likely to peak during  January 2024 (with the range of July 2023 to September 2024), based on its empirically estimated amplitude of \(116.91\pm2.89\). Note that this estimated amplitude matches with the physical model based prediction of \cite{Bhowmik_2018_NatComm}.

Predicting the time of maximum amplitude of sunspot cycle is important for gauging when the most adverse space environmental conditions (space weather) are expected. This information is important for solar radiative forcing of the Earth's upper atmosphere, in protection of space based technological assets and mission lifetime estimates. This prediction of the timing of the peak of sunspot cycles have remained a challenging task for physics based models. We have provided an alternative empirical method for predicting the timing of the sunspot cycle peak which can be implemented only after a significant fraction of the rising phase of sunspot cycle has occurred. The physical model based prediction of \cite{Bhowmik_2018_NatComm} predicted the peak to occur in 2024 (\(\pm 1\) year). This convergence of our empirical prediction with early, physics based prediction augurs well for the field of solar cycle predictions.

\section*{Acknowledgements}
CESSI is funded by IISER Kolkata, Ministry of Education, Government of India. C.S. acknowledges fellowship from CSIR through grant no. 09/921(0334)/2020-EMR-I. The authors acknowledge helpful exchanges during the third team meeting of ISSI Team 474 sponsored by the International Space Science Institute, Bern. Authors are thankful to an anonymous reviewer for constructive comments.

%%%%%%%%%%%%%%%%%%%%%%%%%%%%%%%%%%%%%%%%%%%%%%%%%%
\section*{Data Availability}

We use total sunspot number data made available by WDC-SILSO\footnote{\href{https://www.sidc.be/SILSO/datafiles}{https://www.sidc.be/SILSO/datafiles}}, Royal Observatory of Belgium, Brussels. We also make use of Wilcox Solar Observatory synoptic charts\footnote{\href{http://wso.stanford.edu/synopticl.html}{http://wso.stanford.edu/synopticl.html}}. Scripts of our statistical analyses will be shared on reasonable requests to the corresponding author.

%%%%%%%%%%%%%%%%%%%% REFERENCES %%%%%%%%%%%%%%%%%%

% The best way to enter references is to use BibTeX:

\bibliographystyle{mnras}
\bibliography{references} % if your bibtex file is called example.bib

% Alternatively you could enter them by hand, like this:
% This method is tedious and prone to error if you have lots of references
%\begin{thebibliography}{99}
%\bibitem[\protect\citeauthoryear{Author}{2012}]{Author2012}
%Author A.~N., 2013, Journal of Improbable Astronomy, 1, 1
%\bibitem[\protect\citeauthoryear{Others}{2013}]{Others2013}
%Others S., 2012, Journal of Interesting Stuff, 17, 198
%\end{thebibliography}

%%%%%%%%%%%%%%%%%%%%%%%%%%%%%%%%%%%%%%%%%%%%%%%%%%

%%%%%%%%%%%%%%%%% APPENDICES %%%%%%%%%%%%%%%%%%%%%

\appendix

\section{Determination of decay time interval of dipole moment cycles}
\label{sec: Appendix}

\begin{figure*}
    \centering
    \includegraphics[width = 0.9\textwidth]{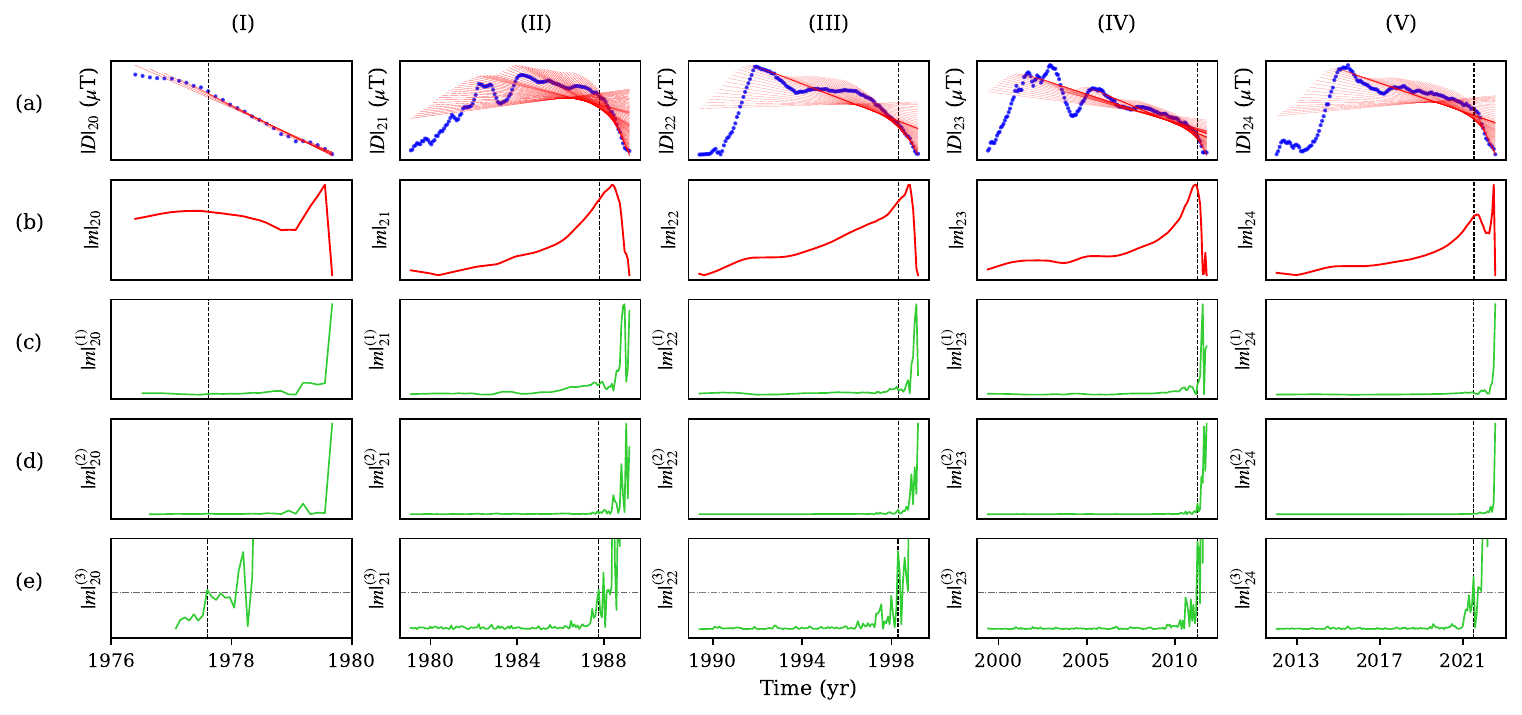}
    \caption{Panel (a): unsigned dipole moment cycles \(\vert{} D\vert{}_{20-24}\) are depicted (in blue) across columns (I)-(V), overlaid by the linear regression fits (in red). Regression fits are implemented starting from each individual data point, one by one, till the end of the respective cycles. Panel (b): time evolution of slopes of linear fits computed in panel (a). Panel (c)-(e): first (\(\vert{}m\vert{}^{(1)}\)), second (\(\vert{}m\vert{}^{(2)}\)) and third order(\(\vert{}m\vert{}^{(3)}\)) forward differences of these slopes are shown along these three rows (c)-(e), respectively. In all the panels, vertical black dashed lines mark the beginning of decay time interval i.e.,  the instant when plateau-like phase in a dipole moment cycle ends, or in other words, the dipole moment cycle starts decaying rapidly with almost a uniform rate. The sudden changes in \(\vert D\vert\) associated with this instant is captured well in the third order forward difference, \(\vert{}m\vert{}^{(3)}\), in the form of pronounced peaks [panel (e)]. To maintain uniformity and robustness of the choice of decay interval across individual \(\vert D \vert\) cycles, we consider the first pronounced peak with amplitude >0.4 in \(\vert{}m\vert{}^{(3)}\) time series to mark the initial epoch of the decay time interval. This interval ends with the termination of the corresponding \(\vert D\vert\) cycle.}
    \label{fig:Figure A1}
\end{figure*}

Fig.\ref{fig:Figure A1} summarizes the technique used to determine the decay intervals of the unsigned dipole moment cycles \(\vert{} D\vert{}_{20-24}\) based on which we calculate their respective rate of decay. For further explanation refer to the caption of the same figure.

\begin{figure}
    \centering    \includegraphics[width=0.45\textwidth]{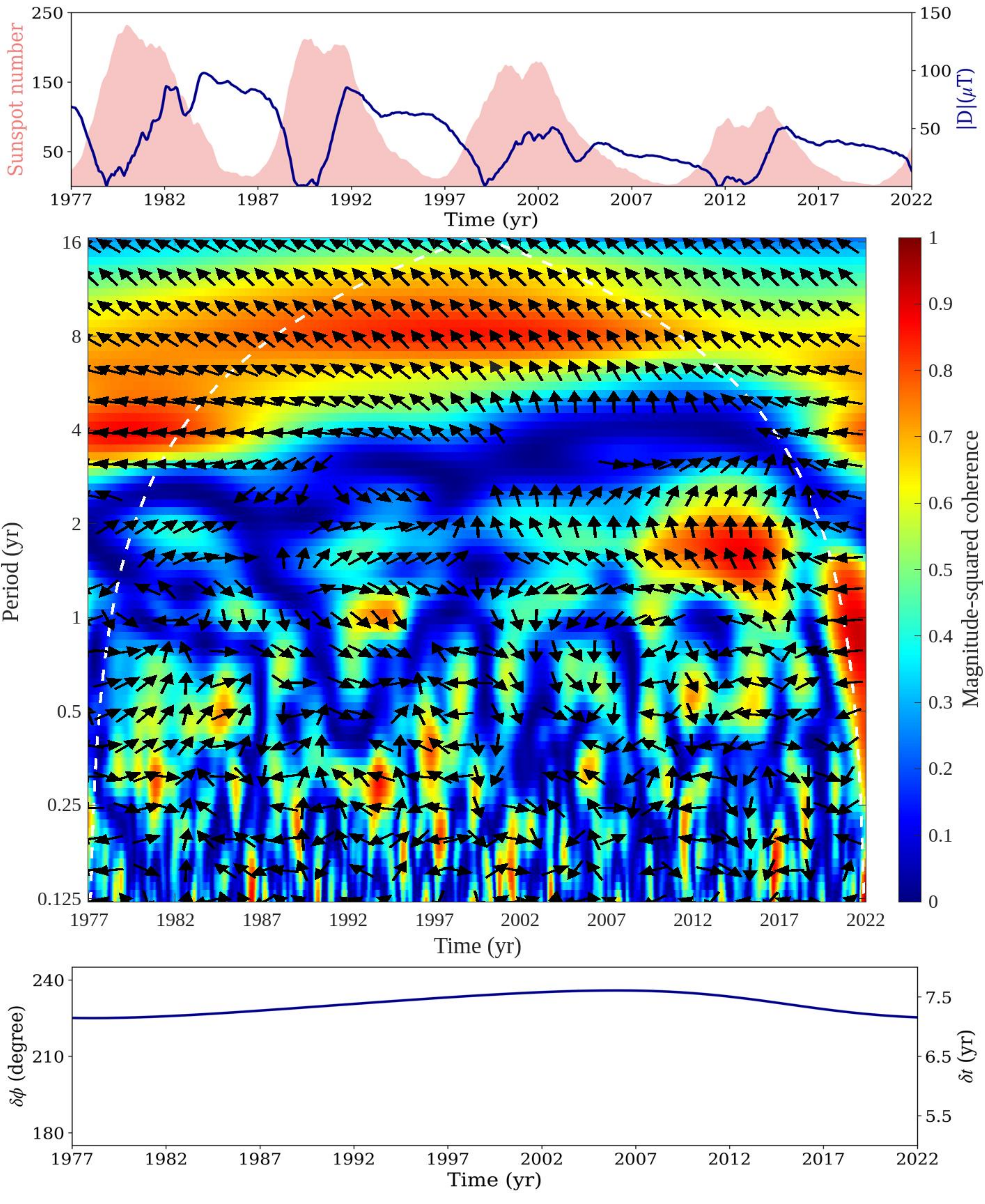}
    \caption{Sunspot number time series and the unsigned dipole moment cycles (top panel) and their wavelet-coherence analysis depicting strong coherence (represented by the color map) and a phase difference (marked by arrows) between them around the 11-yr periodicity (middle panel), wherein, the white-dashed curve denotes the cone-of-influence. Temporal evolution of the phase difference between the two series is shown in the bottom panel.}
    \label{fig:figA2}
\end{figure}

\begin{figure}
    \centering    \includegraphics[width=0.44\textwidth]{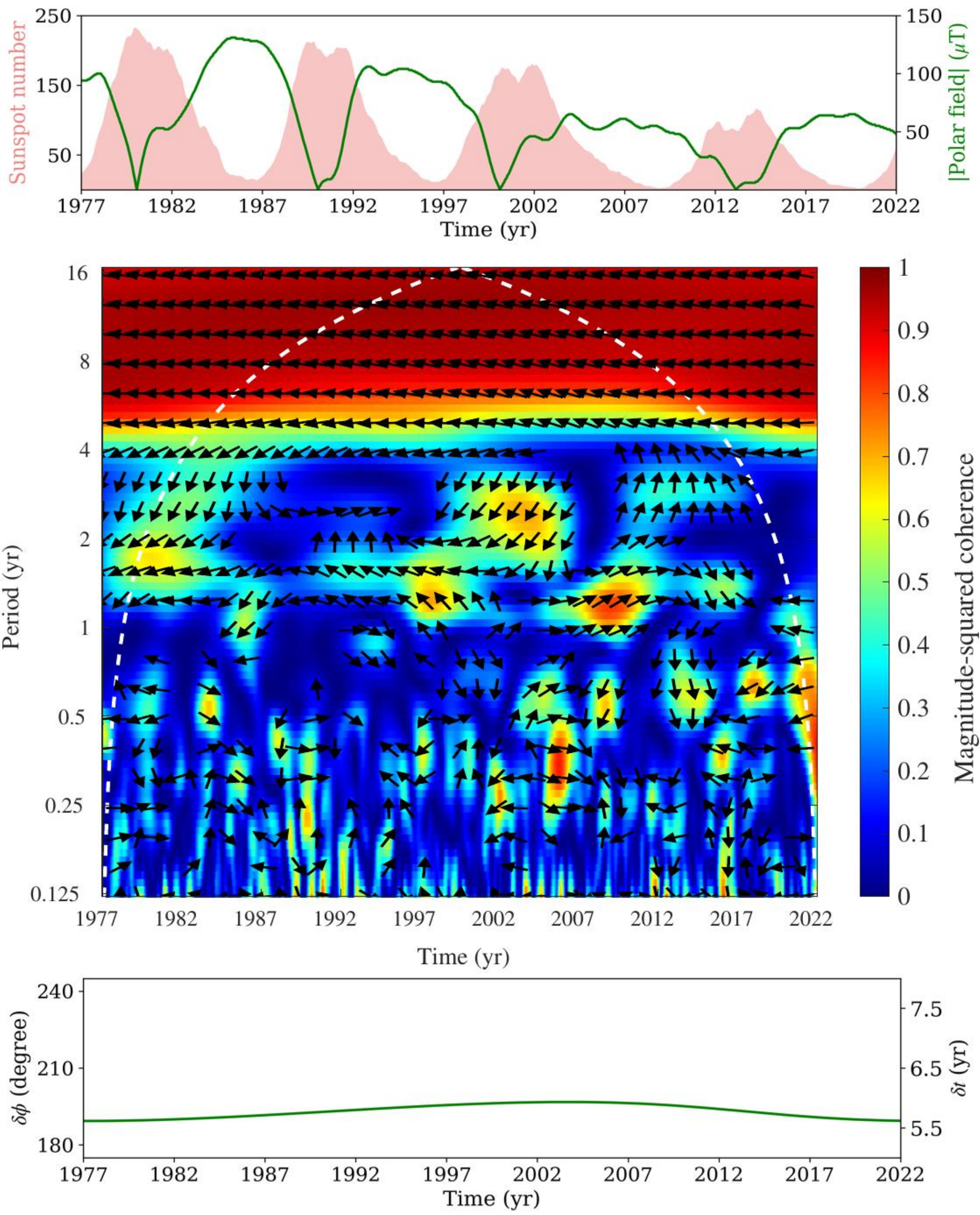}
    \caption{Same as in Fig.\ref{fig:figA2}, but for the wavelet-coherence between the sunspot cycles and the polar field time series.}
    \label{fig:figA3}
\end{figure}

The wavelet-coherence analysis (Ref. \citeauthor{GRINSTED_2004_NPG}  \citeyear{GRINSTED_2004_NPG}) in Fig.\ref{fig:figA2} and Fig.\ref{fig:figA3} show that the unsigned dipole moment and WSO polar field lead the sunspot time series on average by a phase of \(\sim225^{\circ}\) (\(\sim7\) years) and \(\sim180^{\circ}\) (\(\sim5.5\) years), respectively. This clearly indicates that the average polar field decays more concurrently with the growth of the subsequent sunspot cycle.

% Don't change these lines
\bsp	% typesetting comment
\label{lastpage}
\end{document}